\documentclass[preprint,12pt]{elsarticle}

\usepackage{amssymb}
\usepackage{graphicx}
\usepackage{dcolumn}
\usepackage{bm}
\usepackage{subfigure}
\usepackage{amsfonts,amsmath}
\usepackage{color}

\journal{ Physica A}

\begin{document}

\begin{frontmatter}



\title{Abnormal Synchronizing Path of Delay-coupled Chaotic Oscillators on the Edge of Stability}

 \tnotetext[t1]{This document is a draft for Physica A.}

 \author[ustc]{Zhao Zhuo}

 \author[uestc]{Shi-Min Cai\corref{cor1}}
 \ead{ shimin.cai81@gmail.com, csm1981@mail.ustc.edu.cn}

 \author[ustc]{Zhong-Qian Fu}

 \cortext[cor1]{Corresponding author. Tel.: +86 28 61830223.}

\address[ustc]{Department of Electronic Science and
Technology, University of Science and Technology of China, Hefei
Anhui, 230026, P. R. China}
\address[uestc]{Web Sciences Center, School of Computer Science and Engineering
University of Electronic Science and Technology of China,
Chengdu Sichuan, 611731, P. R. China}

\begin{abstract}
In this paper, the transition of synchronizing path of delay-coupled
chaotic oscillators in a scale-free network is highlighted. Mainly, through the
critical transmission delay makes chaotic oscillators
be coupled on the edge of stability, we find that the transition
of synchronizing path is \emph{abnormal}, which is characterized by
the following evidences: (a) synchronization process starts with low-degree
rather than high-degree ones; (b) the high-degree nodes don't undertake the role of
hub; (c) the synchronized subnetworks show a poor small-world property as a result of
hubs absence; (d) the clustering synchronization behavior emerges even
community structure is absent in the scale-free network. This abnormal
synchronizing path suggests that the diverse synchronization behaviors
occur in the same topology, which implies that the relationship between
dynamics and structure of network is much more complicated than the common
sense that the structure is the foundation of dynamics. Moreover,
it also reveals the potential connection from the transition of synchronization
behavior to disorder in real complex networks, e.g. Alzheimer disease
\end{abstract}


\end{frontmatter}


\section{Introduction}

Synchronization, which is referred to as the uniform activities of
interacting units, is ubiquitous in nature and plays an important role in
physics, biology, sociology, technology and neural systems
\cite{Pikovsky2002,Arenas2008}. Generally, real-world systems
are often represented by networks, of which the nodes
are units and edges indicate their interactions \cite{Watts1998,Albert2002,Newman2003,Boccaletti2006}.
A networked system consisting of $N$ coupled oscillators (i.e., nodes) thus can be simply
described by $\dot{\mathbf{x}}_i=\mathbf{F}(\mathbf{x}_i)-c\sum_il_{ij}\mathbf{H}(\mathbf{x}_j)$,
where $\mathbf{x}_i$ describes the the $m$-dimensional state of node $i$ ($i=1,2,...,N$),
$\mathbf{F}$ indicates nodes' dynamics, $c$ and $F$ are the coupling strength and function,
and $\mathbf{L}=(l_{ij})_{N\times N}$ is the Laplacian matrix of network.
It reaches synchronization if all oscillators satisfy
$\mathbf{x}_1(t) \equiv \mathbf{x}_2(t) \equiv ... \equiv \mathbf{x}_N(t)$.

The synchronizability of network usually emphasizes whether its synchronization
is achievable. Pecora and Carroll \cite{Pecora1998} varied coupling strength and firstly unraveled
that the synchronizability of network could be evaluated by the eigenratio $\lambda_{max}/\lambda_2$,
where $\lambda_{max}$ and $\lambda_2$ were the maximum and minimum non-trivial
eigenvalue of $\mathbf{L}$. The smaller the eigenratio is,
the stronger synchronizability the network has, and \emph{vise versa}.
Following this pioneering work, many researches have comprehensive investigated
how to improve synchronizability of network via edge betweenness \cite{Chavez2005},
topology modification \cite{Zhao2005,Yin2006}, optimization \cite{Donetti2005,Nishikawa2006a,Nishikawa2006b},
adaptive evolution \cite{zhou2006}, and so on. Besides the network structure,
Li and Chen \cite{Li2004} found that transmission delay between oscillators also influenced synchronizability of network deeply
due to the signals traveling in limited speed, such as neural oscillators \cite{Dhamala04}.

Moreover, the temporal characteristics shown in synchronization process
can provide more information of the structure and function of system.
For examples, the temporal correlations between interacting neural oscillators
represent the clustering structure of coupling topology \cite{Zhou2006b,Zhuo2011},
and clustering synchronization has been applied in community detection
\cite{Arenas2006,ZMY2014} and data clustering \cite{Bohm2010,Hong2012}. It is also worthy to discuss
one of the most attractive temporal characteristics, synchronizing path that describes
how the synchronized state diffuses to whole network. G\'omez-Garde\~nes \emph{et al}
\cite{Gomez2007} studied the synchronization path based on Kuramoto model, and
showed it behave difference from Barab\'asi-Albert scale-free (BA) network
to Erd\"os-R\'enyi random (ER) network. Especially, the high-degree nodes in
BA network were more likely to be synchronized into the central cluster than
low-degree ones. And, Stout \emph{et al} \cite{Stout2011} proved that the synchronizing
path observed in Ref. \cite{Gomez2007} was robust to increasing oscillator number
and changing the distribution function of the Kuramoto oscillators' natural
frequencies. Recently, Zhou \emph{et.al} \cite{Zhou2013} studied the influence
of the mixing pattern on synchronizing path and found that the disassortative
BA networks synchronize via a merging process rather than the typical growth
process on BA network.

In this paper, we mainly investigate how the transmission delay affects the synchronization
of chaotic oscillators. Based on a BA network consisting of R\"ossler oscillators,
two distinct types of synchronizing paths are found via controlling transmission delay.
Especially, when transmission delay achieves a critical value, the synchronizing path shows abnormal
evidences that (a) these low-degree nodes rather than high-degree ones firstly synchronized,
(b) the high-degree nodes no longer act as hubs in synchronization process
because they join the largest synchronized cluster very late, (c) the largest
synchronized cluster show poor small-world property as a result of absent
hubs, (d) the clustering synchronization behaviors appears even the BA network is in
absence of community. These features make the abnormal synchronizing path on the edge of
stability (i.e., critical transmission delay) obviously differ from that on the stable region,
and also reveal the disagreement between structure and dynamics of network.
Meanwhile, this transition of synchronizing path suggests
that the same topology can display diverse synchronization behaviors,
which may provide a deep insight to understand the Alzheimer disease.

\section{Synchronization of Delay-coupled Chaotic Oscillators}

The network consisting of $N$ delay-coupled oscillators is represented by
\begin{equation}
\dot{\mathbf{x}}_i(t) = \mathbf{F}[\mathbf{x}_i(t)]-c\Sigma^N_{j=1}l_{ij}\mathbf{H}[\mathbf{x}_j(t-\tau)],
\label{equ:system}
\end{equation}
where $\tau$ is the transmission delay, $c$ is the coupling strength, and $l_{ij}$
is the entry of $\mathbf{L}$ at the $i$-th line and $j$-th column. The coupling
function $\mathbf{H(\mathbf{x})}$ is set in the linear form as
$\mathbf{H(\mathbf{x})}=\mathbf{x}$. As proved in Ref. \cite{Yan2009}, there
exists a critical transmission delay $\tau_c$ that the synchronization is stable
(unstable) if $\tau<\tau_c$($\tau>\tau_c$). The synchronization error $W(t)$ is adopted
to measure the degree of synchronization,
\begin{equation}
W(t)= \frac{1}{N}\Sigma_{i=1}^N|\mathbf{x}_i-\langle \mathbf{x}_i\rangle|^2,
\label{equ:Wt}
\end{equation}
where $\langle...\rangle$ denotes averaging over all nodes. Then the logarithmic
synchronizing speed is defined as
\begin{equation}
\mu(t) = -\frac{d\ln[W(t)]}{dt}.
 \label{equ:mut}
\end{equation}
Positive synchronizing speed means that the $W(t)$ vanishes and the synchronization is stable.
Thus, the critical transmission delay can be found by checking the sign of synchronizing speeds.

The R\"ossler oscillator is chosen for the numeric estimation of synchronous behaviors,
which reads as
\begin{equation}
\left\{
\begin{aligned}
& \dot{x}_{i1}=-x_{i2}-x_{i3},\\
& \dot{x}_{i2}=x_{i1}+0.2 x_{i2},\\
& \dot{x}_{i3}=0.2+x_{i3}(x_{i1}-12).\\
\end{aligned}
\right.
\end{equation}
We use the Barab\'asi-Albert (BA) model with node number $N=1000$ and mean degree $\bar{k}= 10$
as the coupling topology and set coupling strength as $c=3$. The synchronization processes are
carried out by fourth-order Ronge-Kutta method with fixed step $\delta t=1.0\times10^{-4}$.

\begin{figure*}
\centering{
\subfigure[]{\includegraphics[width=2.6in]{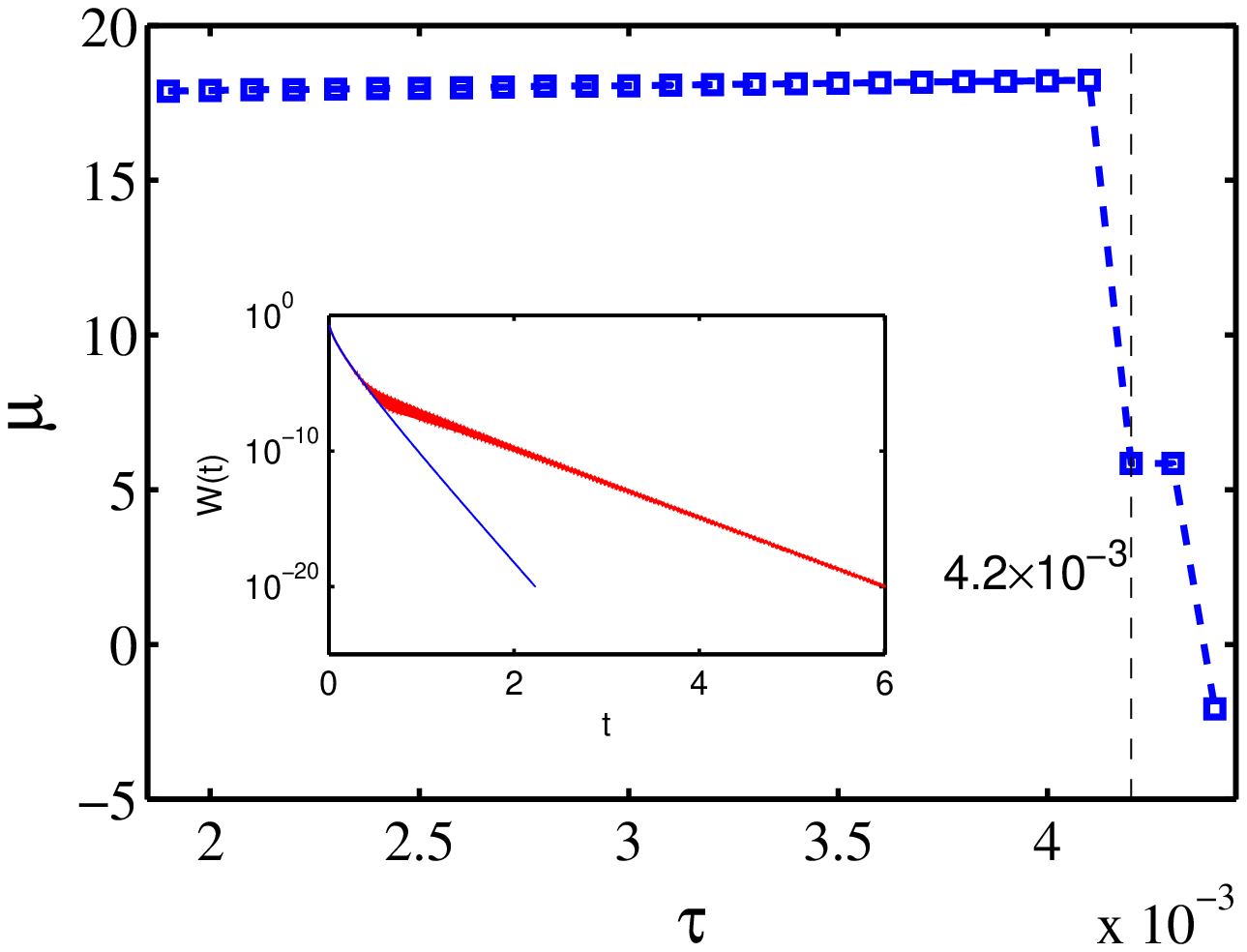}\label{fig:1a}}
\subfigure[]{\includegraphics[width=2.6in]{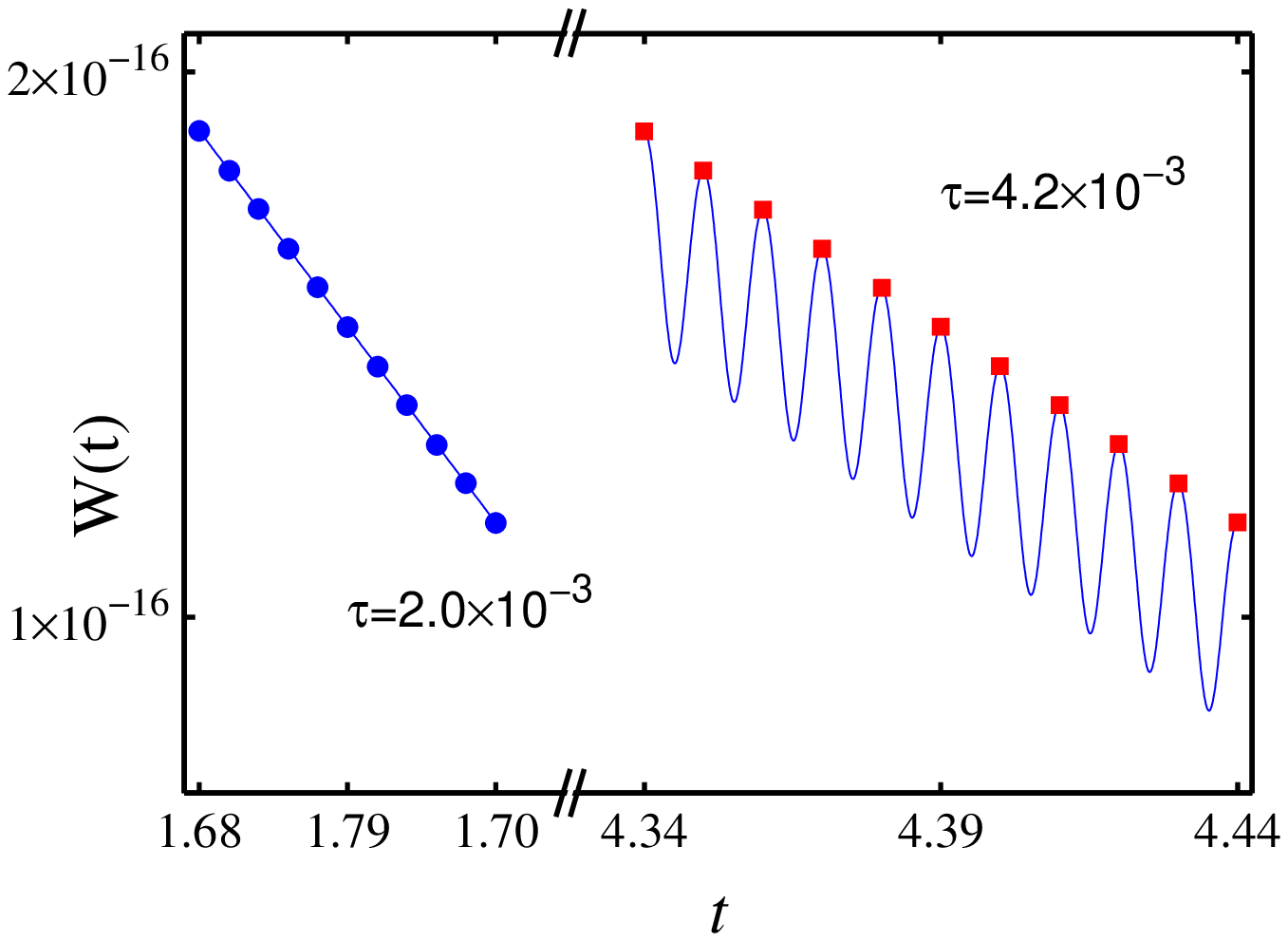}\label{fig:1b}}
\caption{(Color online) (a) The logarithmic synchronizing speed of delay-coupled oscillates.
The critical transmission delay $\tau_c=4.2\times10^{-3}$ is marked by the dashed line. For
the inner plot, it gives temporal evolutions of $W(t)$ when $\tau$ is $2.0\times10^{-3}$
(blue line) and $4.3\times10^{-3}$ (red line). (b) Part of the snapshots of synchronization process.
$\blacksquare$ indicates snapshot of synchronization on the edge of stability and $\bullet$
describes the corresponding one of synchronization on the stable region,
which shows the same synchronization error. Note the break in $x$-axis.}}
\end{figure*}

The synchronizing speeds with varying transmission delays are obtained
by the first-order polynomial fitting of $ln[W(t)]$ after transient
time of $10^4$ steps, which is shown in Fig. \ref{fig:1a}. It
can be seen that the critical transmission delay is around $4.2\times10^{-3}$.
Then the synchronization with $\tau_c=4.2\times10^{-3}$ act as an example
of collective dynamics on the edge of stability and the one
with $\tau=2.0\times10^{-3}$ as that in the stable region
to be compared with. Both of the temporal evolutions of $W(t)$
are shown in the inset plot of Fig .\ref{fig:1a}. In order to
study the temporal synchronization behavior, we capture a
serial snapshots of network state. Since the $W(t)$ for critical transmission delay
is the combination of a decreasing trend and periodic
fluctuation (see Fig. \ref{fig:1b}), the times series $[t_1,t_2,...,t_m]$
are chosen as $W(t_i)$ is the local maximum. For simplicity to compare the temporal synchronization
behaviors, snapshots of synchronization process in the stable region, whose synchronization
degree is indicated by $W'(t)$, are chosen at the time
series $[t'_1,t'_2,...,t'_m]$ satisfying $W(t_i)=W'(t'_i)$.
Only snapshots of $t_i$ after $3\times10^4$ steps are
considered to avoid the disturbance in the transient time.
Total $M=356$ snapshots are selected in each synchronization,
part of which is shown in Fig. \ref{fig:1b}.

\section{Synchronizing Path on the Edge of Stability}

\begin{figure*}
\centering{
\subfigure[]{\includegraphics[width=2.6in]{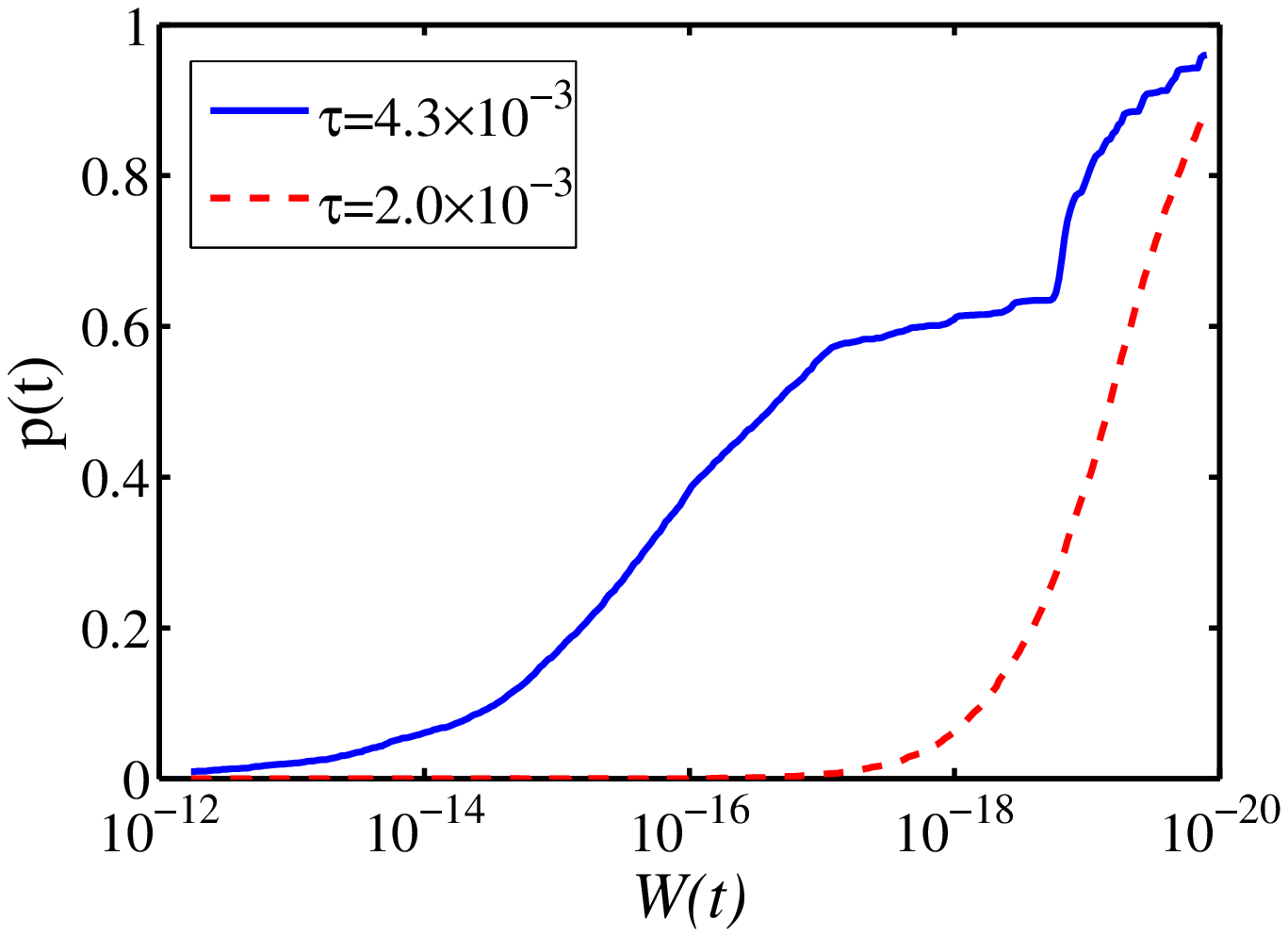}\label{fig:2a}}
\subfigure[]{\includegraphics[width=2.6in]{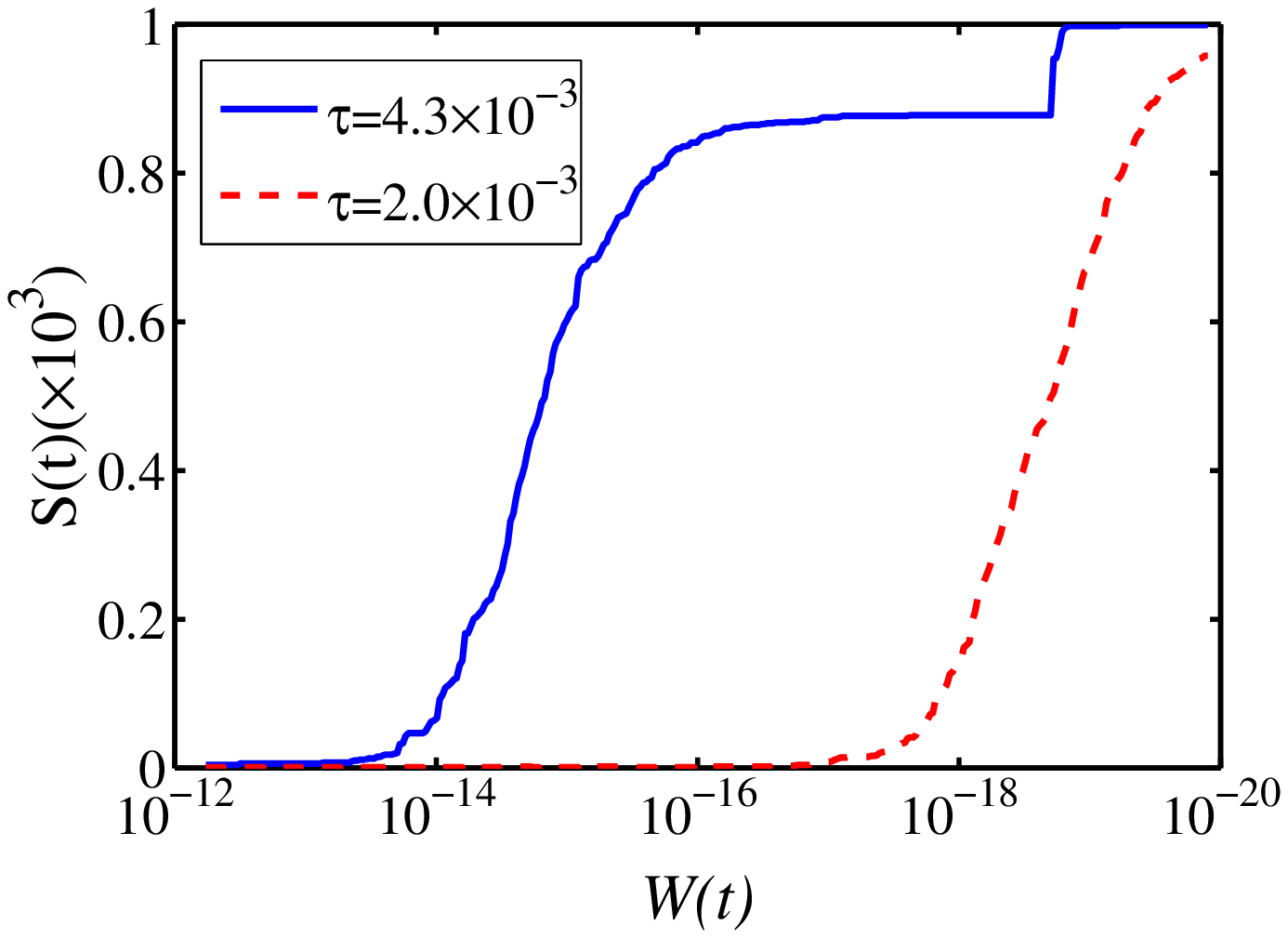}\label{fig:2b}}
\caption{(Color online) (a) The percent of preserved edges. (b) The size of GC.
Both of them imply the difference of synchronizing path between
these two synchronization processes. Note that in both figures, the decreasing order
in $x$-axis indicates the increase of time and decrease of synchronization error.}
}
\end{figure*}

According to the method in Ref. \cite{Gomez2007}, we investigate the synchronizing path via
examining how the synchronized clusters grow in the snapshots.
The synchronized clusters are extracted
as follows. The edge linking node $i$ and $j$ is
weighted by the Euclidean distance
between the nodes
\begin{equation}
w_{\{i,j\}\in E} = |\mathbf{x}_i-\mathbf{x}_j|^2.
\end{equation}
Then a threshold $T$ is set and the edges associated to weights less than the threshold are
preserved to form a new network for each snapshot. In this new networks, two connected nodes
are considered as synchronized. The largest synchronized cluster is defined as the \emph{giant
component} (GC) of the new network. Here we choose the synchronization error of the last snapshot
$W(t_M)=W'(t'_M)$ as the threshold. The percents $p(t)$ of preserved edges
and sizes of GCs $S(t)$ as a function of synchronization error $W(t)$ for
two synchronization processes are show in Fig. \ref{fig:2a} and \ref{fig:2b}, respectively.
Note that the decreasing order in $x$-axis
of both figures is in correspondence to the increase of time and decrease of
synchronization error. One can see that at the same degree of synchronization,
more edges are preserved and the GCs are larger in the snapshots of synchronization with the
critical transmission delay than those in the snapshots of synchronization on the stable region.
This result suggests that at the same degree of synchronization the network is more synchronized
on the edge of stability and imply a different synchronizing path.

\begin{figure*}
\centering{
\subfigure[]{\includegraphics[width=2.5in]{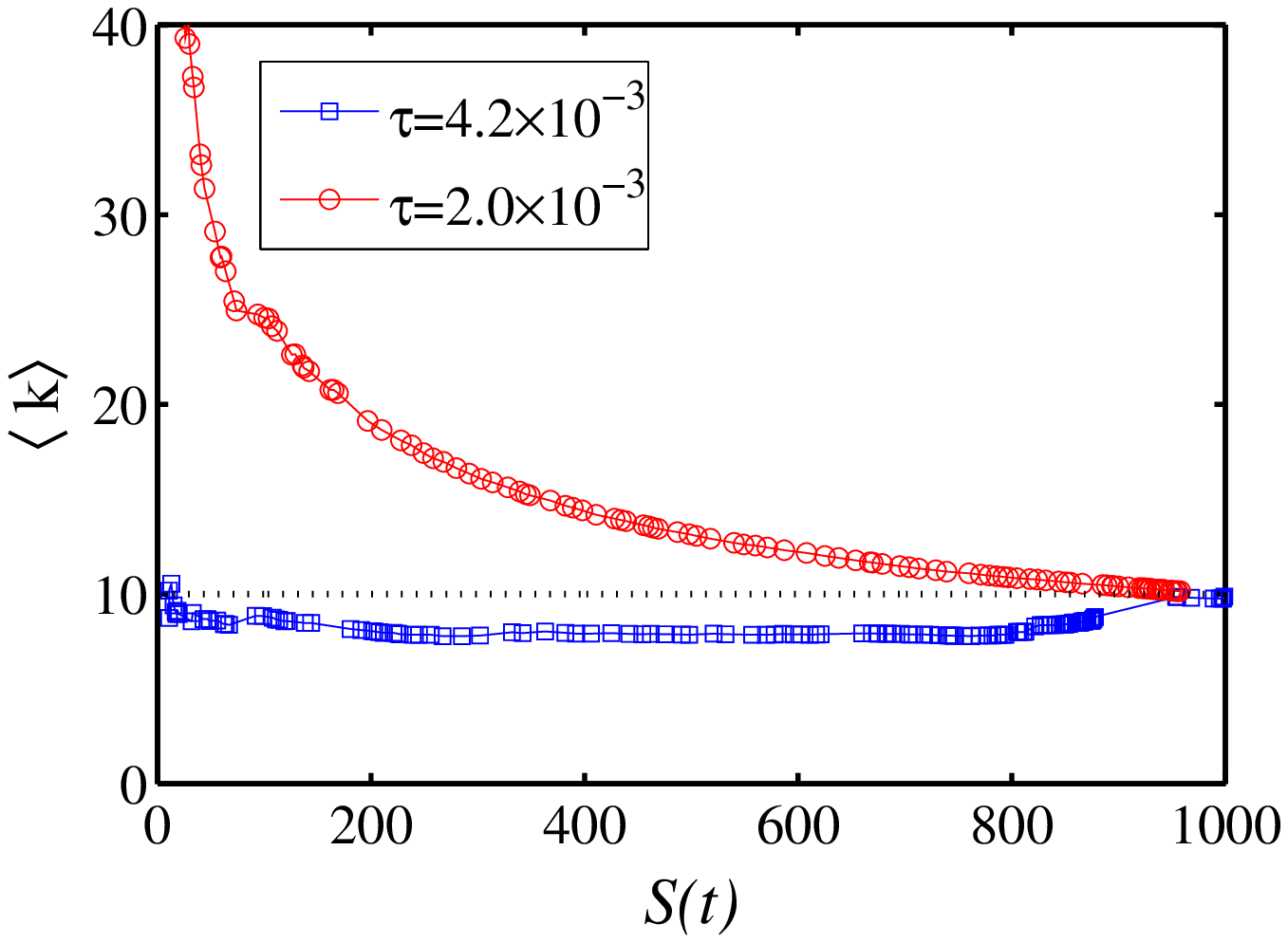}\label{fig:3a}}
\subfigure[]{\includegraphics[width=2.5in]{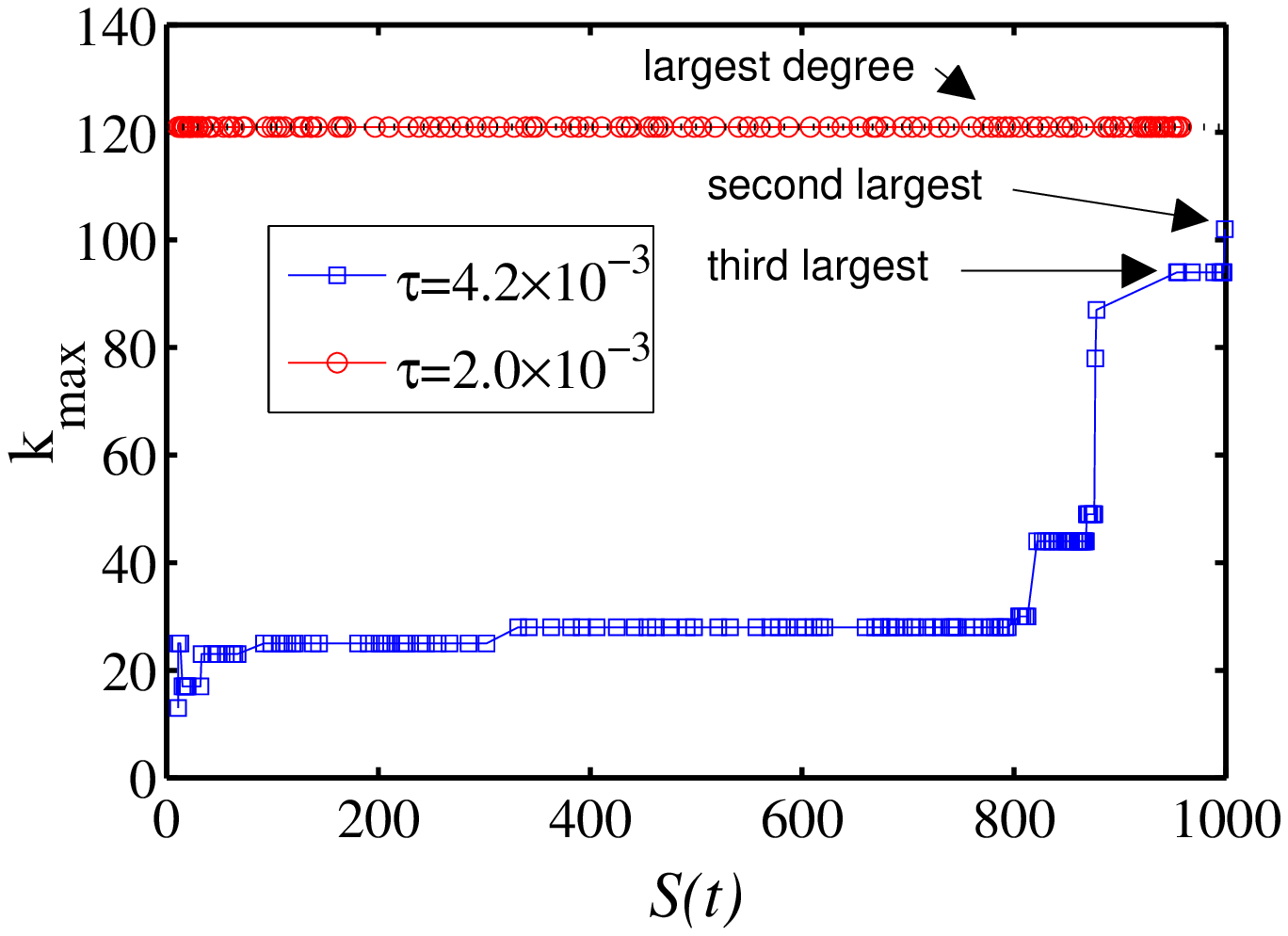}\label{fig:3b}}
\subfigure[]{\includegraphics[width=2.5in]{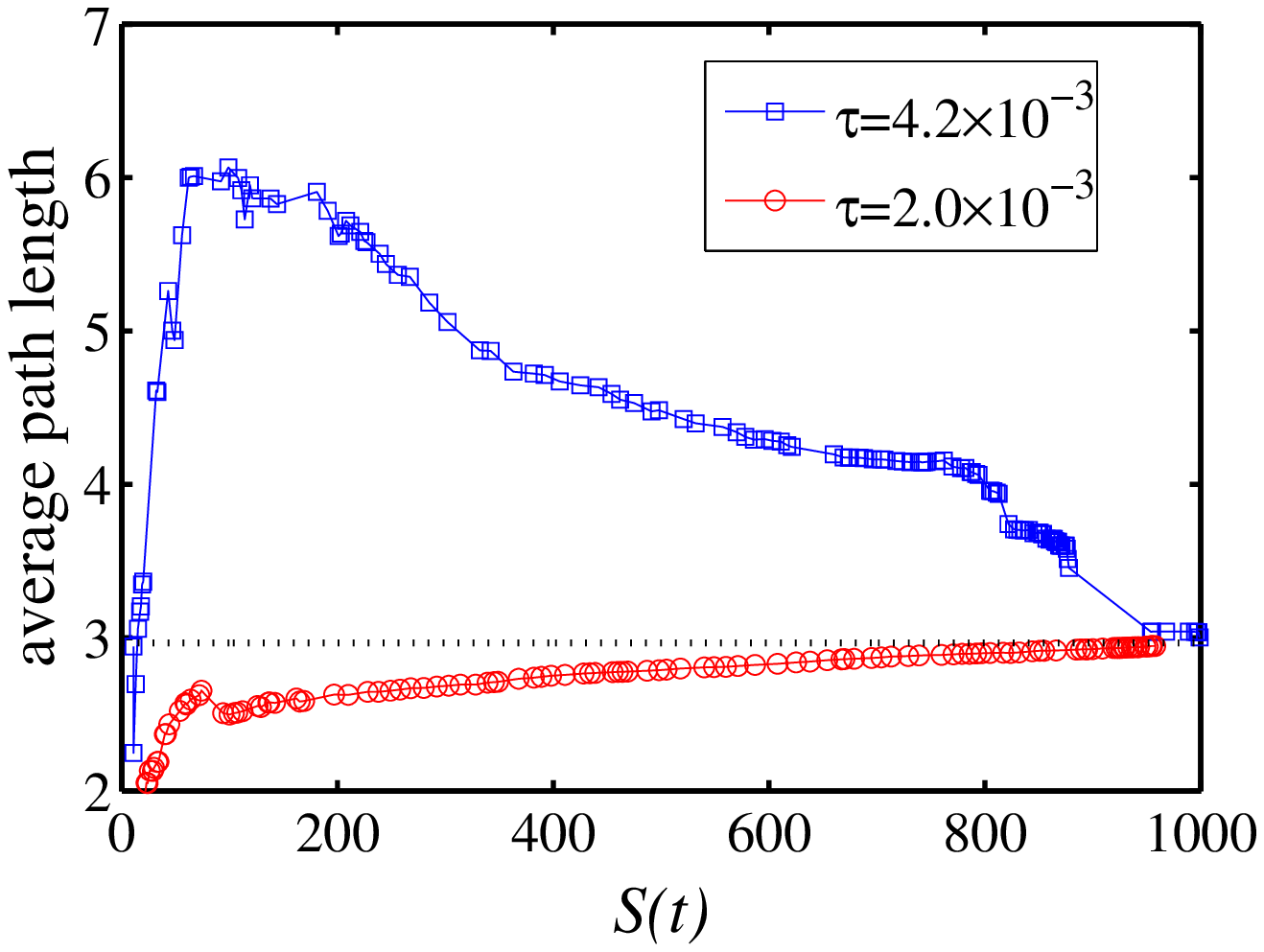}\label{fig:3c}}
\subfigure[]{\includegraphics[width=2.5in]{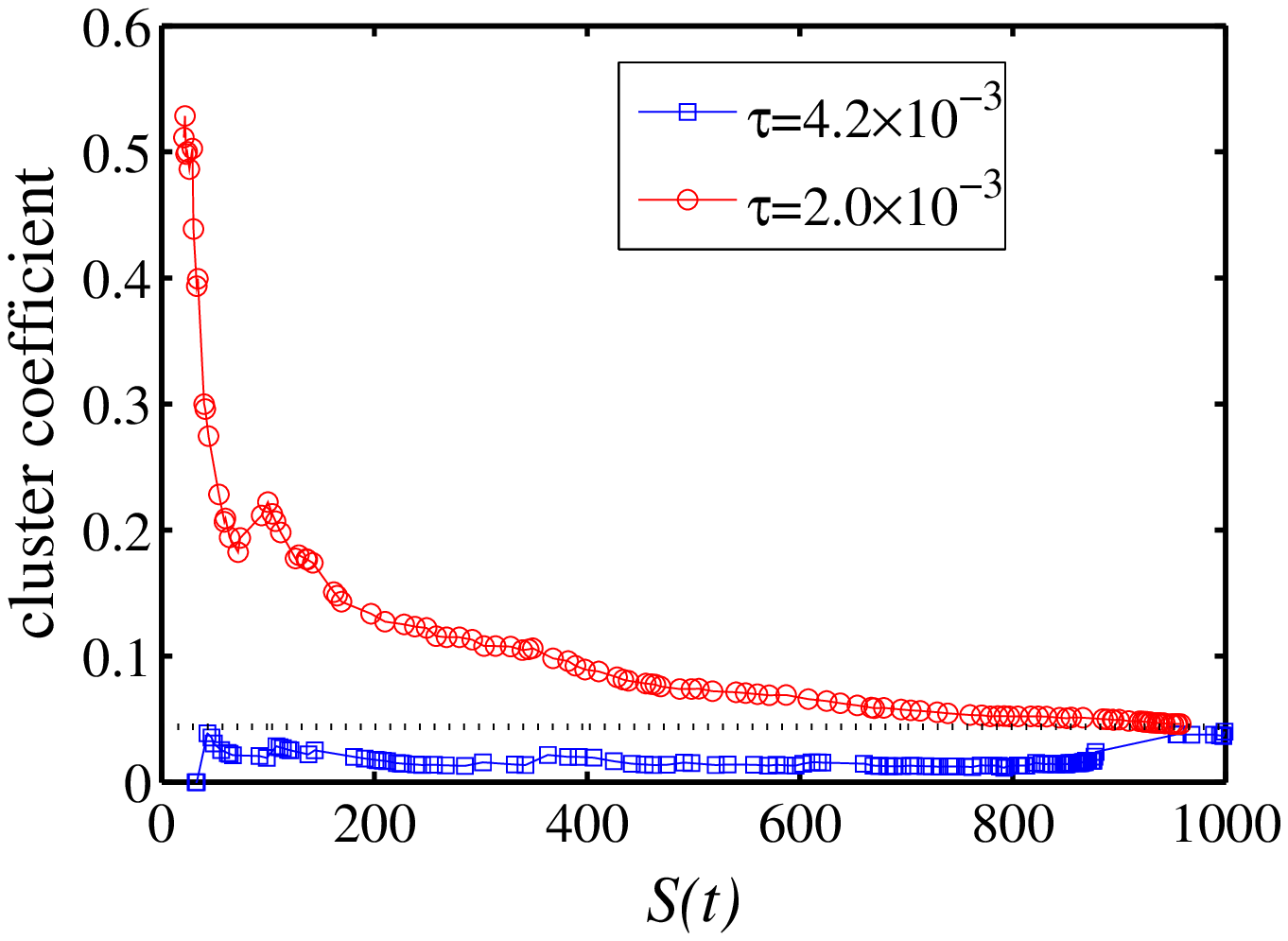}\label{fig:3d}}
\caption{(Color online) (a) The mean degree $\langle k\rangle$
of nodes in the GC, (b) The largest degree $k_{max}$ of nodes
in the GC, (c) the average path length of GC, and (d) the cluster coefficient
of GC as function of its size $S(t)$. The dotted line
indicates the mean degree, largest degree, average path length and cluster coefficient of network
in (a), (b), (c) and (d), respectively. Note that in all figures, the increasing order
in $x$-axis indicates the increase of time and decrease of synchronization error.}
}
\end{figure*}

It has been observed that nodes' degrees are tightly connected to the synchronization process, e.g.
phase synchronization starts at high-degree nodes \cite{Gomez2007} and explosive synchronization
emergences if the natural frequencies of oscillators depend on their degrees \cite{Leyva2012,Li2013}.
Naturally, we examine the degrees (corresponding to BA network) of nodes in the GC to give a details of synchronizing
path. In Fig. \ref{fig:3a}, it shows the mean degree $\langle k\rangle$ of nodes in the GC
as function of its size $S(t)$. The referred baseline is the mean degree $\bar{k}=10$ of the
BA network. Concretely, when the size of GC is small, the mean degrees of the synchronization
on the stable region are much higher than $\bar{k}=10$, which suggests that
the synchronization starts at the high-degree nodes. As the size of GC grows, the
the mean degree decreases to $\bar{k}=10$ because of the
join of low-degree nodes. on the contrary, the synchronization on the edge of stability starts
with some low-degree nodes since the mean degree is around $\bar{k}$ at the beginning,
then the mean degree further decreases to smaller than $\bar{k}$ because more low-degree nodes join.
It starts to increase till the GC is relevantly large which implies that the high-degree nodes
join the GC very late. To further determine the result, we present the largest degree $k_{max}$ in the GC
as a function of $S(t)$, shown in Fig. \ref{fig:3b}. It can be seen
that the largest degree $K_{max} = 121$ of the BA network join the GC at the very beginning for
transmission delay $\tau=2.0\times10^{-3}$, while it is isolated for critical transmission delay.
In the BA network, high-degree nodes act as structural hubs and play an
important role in dynamic processes such as pinning control \cite{Wang2002}
and synchronization \cite{Pereira2010}. However, they lost their roles as hubs in the
abnormal synchronizing path, which suggests the disagreement between structure
and dynamics of network. As it's unambiguously agreed that many networks realize their functions via
dynamical processes, the disagreement will lead to disorder of system.

As a result of the hubs' absence, the GCs in the synchronization
on the edge of stability display poor small-world properties.
As shown in Figs.\ref{fig:3c} and \ref{fig:3d}, it is
characterized by the larger average path lengths and lower cluster
coefficients compared with their baselines of the BA network.
However, for the synchronization on the stable region,
the average path lengths are shorter and cluster coefficients are higher
in the normal synchronizing path. This variation of
small-world property is also observed in the collective dynamics of real
complex systems, e.g. the brain, which usually suggests disorders in the neural activities. In an
study based on fMRI of Alzheimer disease (AD), clustering was significantly reduced in brain
functional networks and the loss of small-world property is able to discriminate AD patients from
age-matched comparison subjects with high specificity and sensitivity \cite{Supekar2009}. In
another EEG study of AD, path lengths in brain functional networks constructed by $\beta$-band
($15-35$ Hz) waves were significantly increased in AD patients \cite{Stam2007}. Therefore, the
observation of small-world properties further suggests the abnormal synchronizing path is potentially
related to disorders in real complex systems.

\begin{figure*}
\centering{
\subfigure[]{\includegraphics[width=2.5in]{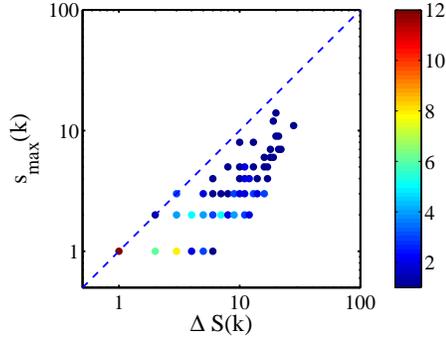}\label{fig:4a}}
\subfigure[]{\includegraphics[width=2.5in]{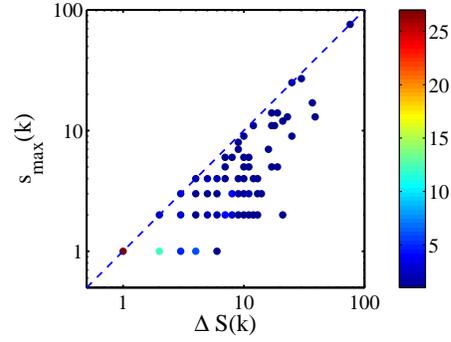}\label{fig:4b}}
\subfigure[]{\includegraphics[width=2.5in]{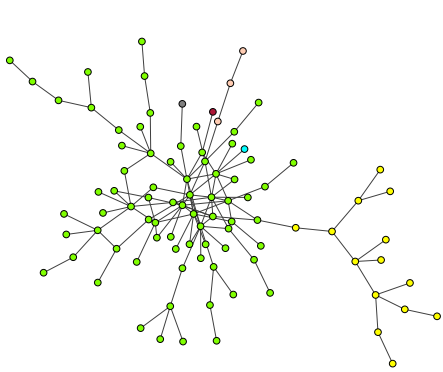}\label{fig:4c}}
\subfigure[]{\includegraphics[width=2.7in]{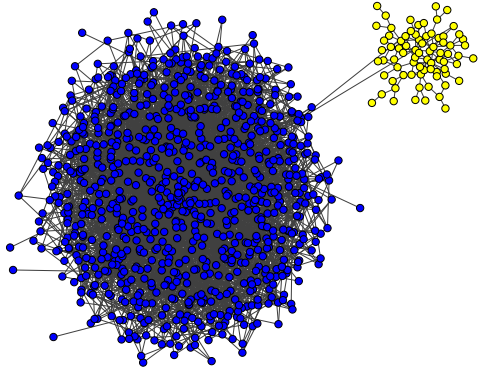}\label{fig:4d}}
\caption{(Color online) (a) $\Delta S(k)$ vs. $s_{max}$ of normal synchronizing path
(b) $\Delta S(k)$ vs. $s_{max}(k)$ of abnormal synchronizing path (c) the growth of GC
in normal synchronizing path, $\Delta S(k)=20$, $s_{max}(k)=14$ and $k=259$.
The size of GC increases from $74$ to $94$. (d) the growth of GC in abnormal synchronizing path,
$\Delta S(k)=s_{max}(k)=76$ and $k=301$. The size of GC increases from $878$ to $954$.}
}
\end{figure*}

To investigate how the GCs diffuse to the whole network in synchronization
with transmission delay, we examine whether the nodes newly joining the GC at each snapshot are
isolated ones or in synchronized clusters at the previous snapshot. Assuming $V(k)$ is the
set of the nodes newly joining the GC at the $k$-th snapshot and $V'(k,c_i)$ is the subset
of $V(k)$ containing the nodes that belong to synchronized clusters labeled
by $c_i$ at the $(k-1)$th snapshot. Denote the number of newly joined nodes as $\Delta S(k)=|V(k)|$
and the maximum size of subsets as $s_{max}(k) = \max_i|V'(k,c_i)|$.
The points of ($\Delta S(k)$,$s_{max}(k)$) of two synchronizing paths are shown in
Fig. \ref{fig:4a} and \ref{fig:4b} respectively. Color of points indicates their number of
occurrence. Small ($\Delta S(k)$, $s_{max}(k)$) pairs represent the nodes connect
to GC as isolated ones, which takes most parts of ($\Delta S(k)$, $s_{max}(k)$) pairs in
normal synchronizing path (see Fig. \ref{fig:4a}).
High $\Delta S(k)$ and $\Delta S(k)\approx s_{max}(k)$ imply the nodes join
the GC as one synchronized cluster, which can be found in the abnormal synchronizing path (see
Fig. \ref{fig:4b}). Fig. \ref{fig:4c} and  \ref{fig:4d} respectively show two examples of GC's growth
, whose $\Delta S(k)$ and $s_{max}(k)$ are both large, in normal and abnormal synchronizing paths
via drawing network using Pajek \cite{pajek}. In both figures, the separated clusters
in the previous snapshot are indicated by different colors. The clustering synchronization
behavior is clearly demonstrated in Fig. \ref{fig:4d}, though
the communities don't exist in the BA network. The result further
supports the disagreement between structure and dynamics in abnormal synchronizing path.

\section{Conclusion}

In summary, we have shown that transmission delay can deeply influence the temporal synchronization
behavior. Even the synchronization processes are all exponential convergence on the edge of stability
and on the stable region, they display completely different synchronizing paths.
In the synchronization with critical transmission delay, the high-degree nodes loss their roles
as hubs in the collective dynamics. As a result, the synchronized part of network displays poor
small-world properties in the abnormal synchronizing path. In further, a detailed investigation
of the growth of the largest synchronized clusters suggests that the clustering synchronization behaviors
are contained in abnormal synchronizing path though the communities don't exist in BA model.
Our work demonstrates that the same network structure can display different dynamic behaviors
with varying transmission delay, and the transition of synchronizing path leads to
disagreement between structure and dynamics, which is probably related to the disorder
of real complex systems, e.g. the disease of human brain.

\section{Acknowledgements}
This work is jointly supported by the National Nature Science Foundation of China (Nos. 60974079 and
61004102), China Postdoctoral Science Foundation (No.
2013M541840), and the Fundamental Research Funds
for the Central Universities (No. ZYGX2012J075)





\end{document}